\documentclass[%
 reprint,
 amsmath,amssymb,
 aps,
]{revtex4-2}

\usepackage{graphicx}
\usepackage{dcolumn}
\usepackage{bm}

\begin{document}

\preprint{APS/123-QED}

\title{Experiments and simulations demonstrating the rapid ultrasonic rewarming of frozen beef cryovials}

\author{Rui Xu}
 \email{rui.xu@ucl.ac.uk}
\author{Bradley E Treeby}%
\author{Eleanor Martin}
\affiliation{Department of Medical Physics and Biomedical Engineering, University College London, London, UK}%

\date{\today}

\begin{abstract}
The development of methods to safely rewarm large volume cryopreserved biological samples remains a barrier to the widespread adoption of cryopreservation.
    Here, experiments and simulations were performed to demonstrate that ultrasound can increase rewarming rates relative to thermal conduction alone.
    An ultrasonic rewarming setup based on a custom 444\,kHz tubular piezoelectric transducer was designed, characterized, and tested with 2\,mL cryovials filled with frozen ground beef.
    Rewarming rates were characterized in the -20$^{\circ}$C to 5$^{\circ}$C range.
    Thermal conduction-based rewarming was compared to thermal conduction plus ultrasonic rewarming, demonstrating a ten-fold increase in rewarming rate when ultrasound was applied.
    The maximum recorded rewarming rate with ultrasound was 57$^{\circ}$C per minute, approximately 2.5 times faster than with thermal conduction alone.
    Coupled acoustic and thermal simulations were developed and showed good agreement with the heating rates demonstrated experimentally and were also used to demonstrate spatial heating distributions with small ($<3^{\circ}$C) temperature differentials throughout the sample when the sample was below 0$^{\circ}$C.
    The experiments and simulations performed in this work demonstrate the potential for ultrasound as a rewarming method for cryopreserved tissues, as faster rewarming rates may improve the viability of cryopreserved tissues and reduce the time needed for cells to regain normal function.
\end{abstract}

\maketitle


\section{Introduction}

There is an unmet need for effective cryopreservation methods that can be implemented in cellular therapies and for transplantable tissue sections\cite{giwa2017promise,urbani2018multi}, in part due to challenges in rewarming these materials without causing damage. 
Cryopreservation of biological materials may be implemented whether via vitrification or by slow freezing.
Vitrification is the cooling of a medium into a disordered and non-crystalline state in the added presence of cryoprotective agents.
Vitrification reduces cellular damage by reducing the likelihood of ice crystal nucleation and growth\cite{fahy1987biological,fahy2015principles}. 
Ice crystal nucleation is more probable at low sub-zero temperatures while ice crystal growth is more probable at higher sub-zero temperatures.
This presents a problem for rewarming a vitrified medium, as any nucleated ice formed during cooling will grow as the medium is rewarmed\cite{fahy1987biological}. 
In consequence, rewarming rates that preserve vitrified cell viability generally exceed cooling rates by at least an order of magnitude\cite{fahy2015principles}.
An alternative to vitrification is slow freezing\cite{kilbride2014scale,urbani2017long} in which the medium is frozen slowly at a controlled and optimized rate and in the presence of ice nucleating and cryoprotectant agents\cite{massie2014gmp,kilbride2014scale}.  
The medium can then be revived with rewarming rates as low as 1$^{\circ}$C per minute\cite{kilbride2017cryopreservation} depending on the rate of freezing.
However, faster rewarming may improve important outcomes such as cell viability, the restoration of cellular lactate production, and the reduction of bulk thermomechanical stresses if the volume is rewarmed homogeneously \cite{kilbride2017cryopreservation}.
Furthermore, faster rewarming may reduce the time needed post-rewarming to regain normal cellular function \cite{kilbride2017cryopreservation}.

The most common rewarming method is immersion in a warm water bath, which can effectively generate fast rewarming rates (up to 200$^{\circ}$C/min for volumes of the order of 100\,$\mu$L\cite{pegg1984effect}.  
However, large ($\geq$3\,mL) volume rewarming cannot be performed effectively with a warm water bath due to the inherently low thermal conduction within biological media, along with the large thermal mass and thermal lag of a large cryopreserved volume \cite{taylor2019new,kilbride2016spatial}. 
Furthermore, thermal gradients generate mechanical stresses in the medium that can cause gross damage to a cryopreserved organ or tissue\cite{fahy1990physical,rubinsky1980thermal}. 
Alternative rewarming methods that don't rely on thermal conduction have been investigated.
Electromagnetic (dielectric) methods can generate fast rewarming of the order of 100$^{\circ}$\,C/min in larger sub-radiofrequency wavelength spherical and ellipsoidal volumes\cite{robinson2002electromagnetic}.
However, there is a positive relationship between most dielectric constants and temperature, making it difficult to control the heat distribution resulting in runaway heating effects\cite{lu2000combined,wusteman2004vitrification}.
An alternative electromagnetic heating method uses alternating magnetic fields to excite a ferrofluid\cite{rosensweig2002heating}.
Magnetic nanoparticles can be perfused into a volume without significantly changing the medium cooling rate\cite{etheridge2014rf}.
The medium can then be heated at a rate at least equal to dielectric heating, but with less potential for thermal runaway or the generation of large thermal gradients\cite{etheridge2014rf,manuchehrabadi2017improved,wang2015theoretical,solanki2017thermo}.
This method has been applied to vitrified rat and rabbit kidneys, and rat hearts to generate mean heating rates above 50$^{\circ}$C/min without significant thermomechanical damage or ice crystal formation \cite{sharma2021vitrification,gao2022vitrification}.
However, scaling inductive heating to larger human-scaled volumes will result in an increase in heat deposition away from the center of the heating coil due to the formation of eddy currents \cite{sharma2021vitrification}. 
This may present a challenge for obtaining a uniform heating distribution.
An additional challenge with inductive heating is obtaining a homogeneous distribution of magnetic nanoparticles, particularly in organs that are less homogeneously vascularized than a kidney. 
Another new method for rewarming cryopreserved materials is laser gold nanowarming\cite{khosla2018characterization}, which has been used to rewarm vitrified millimetre-sized droplets containing biological media and can maintain high cellular viability\cite{khosla2018characterization,zhan2021cryopreservation,zhan2022pancreatic}.
However, scaling this method past 2D scaffolds\cite{zhan2021cryopreservation} to 3D volumes remains a challenge and there remains a need for a rewarming method with a good penetration depth which minimises hot spots.

Here, ultrasound is explored as a potential rewarming technique.
Ultrasound has long been used to heat normothermic tissues, with many hyperthermia and high intensity focused ultrasound applications either in clinical trials or in use as approved medical procedures\cite{kennedy2003high,izadifar2020introduction}. 
Ultrasound in the sub-MHz to low-MHz range has a high penetration depth in soft tissues at normal physiological temperatures, making it suited to heating the center of large volumes.
Ultrasound can be focused electronically using transducer arrays or geometrically\cite{o1949theory} by shaping the source or using lenses or acoustic holograms\cite{lalonde1993field}. 
Electronic focusing and steering of array transducers can additionally be employed to shape or move the focal region over a large volume\cite{do1981annular,hindley2004mri,kohler2009volumetric}. 
Ultrasound may be better suited to rewarming than electromagnetic methods because ultrasound attenuation (which is proportional to the volume rate of heat deposition) decreases as a biological tissue rewarms, reducing the likelihood of thermal runaway\cite{shore1986attenuation}.
Ultrasonic attenuation results from scattering and absorption, the latter of which generates heating\cite{cobbold2006foundations}.
Heating rates in normothermic biological tissues are well characterized and implemented in high intensity focused ultrasound treatments for many conditions including uterine fibroids, pancreatic cancer, and prostate cancer \cite{izadifar2020introduction}, although the contributions of absorption and scattering to attenuation in these tissues is not fully characterised.
It is not known for frozen tissues what proportion of ultrasound is absorbed rather than scattered and this will require investigation.
An experimental demonstration of the rewarming of a frozen tissue sample is needed to demonstrate the viability of ultrasound as a method to rapidly rewarming cryopreserved tissues.

Sound has previously been used to thaw frozen meat and fish, although these experiments were often done in water baths and at sub-100\,kHz frequencies\cite{rosenberg1974method,kissam1982water,guo2021ultrasound}.
In this early work, most of the energy supplied to thaw the materials was supplied by the water bath, with the ultrasound waves assisting in the heat transfer into the materials, without significant energy deposition from the ultrasound itself.
Similar ultrasound-assisted systems are used as freezers, and it has been shown that intermittent ultrasound can increase freezing rates\cite{li2002effect} and increase ice nucleation during freezing\cite{zhang2018using}. 
It has also been shown that 0.4 - 1\,MHz ultrasound can directly thaw frozen meat and fish, using an experimental design where the frozen medium was insulated and not subject to thermal conduction from the surrounding medium\cite{miles1999high}.
However, the thawing times from -10$^{\circ}$C were between tens of minutes and hours depending on the distance from a piston source, and with a close-to linear increasing relationship between thawing times and the distance from the source \cite{miles1999high}.
A common trait of these prior works is that the ultrasonic sources were unfocused, resulting in increased energy deposition in the near field, and focused sources were not investigated. 
A simulation study used multiple sources focused to an ovary to simulate fast (100$^{\circ}$\,C/min) rewarming from low sub-zero temperatures to high sub-zero temperatures\cite{olmo2020use}.
However, simulations did not model rewarming above -20$^{\circ}$C, where the acoustic and thermal properties of biological media have a strong temperature dependence, and experiments are required to confirm the simulated rewarming rates.

In this work, focused ultrasound at 444\,kHz was used to rapidly rewarm biological samples from -20$^{\circ}$C to above 5$^{\circ}$C, and ultrasonic rewarming rates were measured for comparison against other rewarming methods.
The ultrasonic rewarming setup was developed specifically to rewarm biological samples in 2\,mL cryovials. 
Cryovials are commonly used for the cryopreservation of samples (e.g., sperm\cite{li2014ivf}, embryonic stem cells\cite{nagy2006freezing}, and mouse ovaries\cite{candy1997effect}).
The cryovial rewarming experiments were supported by additional experiments that reduce the rewarming effect of thermal conduction from the rewarming setup, demonstrating rewarming primarily caused by ultrasound absorption.

\section{Methods}

\subsection{Design Criteria}

A custom ultrasonic rewarming setup was designed and built.
The design process for the ultrasonic rewarming setup began with three criteria that were intended to ensure that the rewarming setup achieved the rewarming rates necessary for rewarming a slow-cooled material.
The rewarming setup should:\\
1. Repeatably deliver ultrasound to a frozen, filled cryovial without excessive reflective losses.\\
2. Generate a radially centered heat distribution within the cryovial, with limited thermal conduction-based warming.\\
3. Maximize the rewarming rate while maintaining a rewarming volume of sufficient size.\\
The first criterion requires that the ultrasound source be acoustically coupled to the cryovial via a medium that minimizes the impedance mismatch with the frozen biological material stored in the cryovial. 
Olmo $et$ $al.$ used ice as the coupling medium in their simulation study\cite{olmo2020use}, but ice may cause expansion damage during freezing and may decouple during rewarming. 
Natural nylon 6.6 (Direct Plastics UK Ltd.) was chosen instead as the coupling medium, as it has a similar acoustic impedance to frozen beef (See Table \ref{tbl:MediumProps}), reducing reflective losses, and a relatively low acoustic attenuation compared to other plastics.  

\begin{figure}
    \centering
    \includegraphics[width = \columnwidth]{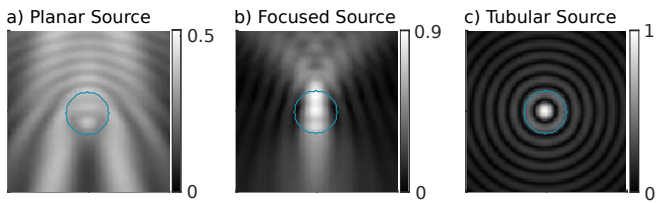}
    \caption{Pressure fields in 12.5\,mm diameter cryovials placed at the a) last axial maximum of a 444\,kHz planar source, b) geometric focus of a 444\,kHz focused source, and c) center of a 444\,kHz tubular source. Pressure fields normalized by c).}
    \label{fig:DesignSims}
\end{figure}

A prospective simulation study was performed to obtain a source geometry satisfying the second criterion, and a source frequency satisfying the third criterion.
Three source geometries were investigated: planar sources (cryovial placed at the location of the last axial maximum), focused arc sources (cryovial placed at the geometric focus), and tubular sources (cryovial placed at the center). 
For simplicity, only single-source setups were considered. 
Figure \ref{fig:DesignSims} shows examples of the pressure magnitude fields of the three types of sources (equal source powers, 444\,kHz) embedded in nylon 6.6, continuously sonicating a cryovial containing frozen beef.

Figure \ref{fig:DesignSims} shows that the tubular source generates the highest pressure, and the pressure distribution is radially centered.
The tubular source is advantaged by its matched geometry to the cryovial.
The frequency displayed in Fig. \ref{fig:DesignSims} shows a central focus and an outer ring encapsulated within the cryovial.
Halving the frequency would result in a central focus covering the entire cryovial, but would likely more-than halve the rewarming rate due to the frequency-dependence of ultrasonic attenuation and corresponding heat deposition. 
In this work the decision was made to use a higher frequency source (400+ kHz as shown in Fig. \ref{fig:DesignSims}) to maximize the rewarming rate at the center of the cryovial.

\subsection{Device Construction}

A tubular modified lead zirconate-lead titanate transducer element was purchased from PI Ceramic (Lederhose, Germany).
The transducer element was made from a proprietary ceramic (PIC181), suited to high power acoustic applications and applications in resonance mode \cite{piceramic}. 
The element was coated with a fired silver electrode (thick film, typically 10\,$\mu$m) covering both inner and outer surfaces, and was radially polarized with the +pole on the outer radial surface.
The outer diameter of the element was 76\,mm, the inner diameter 66\,mm, and the height was 50\,mm.
The radial resonant frequency of the transducer element was first approximated by:
\begin{equation}\label{eq:freq}
    f = \dfrac{N_t}{TH},
\end{equation}
where $TH$ is the thickness of the transducer element (5\,mm) and $N_t$ is the acousto-mechanical frequency coefficient in Hz$\cdot$m (2110\,Hz$\cdot$m for PIC181\cite{piceramic}).
Equation \ref{eq:freq} estimates a resonant frequency of 422\,kHz, but the accuracy of Eq. \ref{eq:freq} decreases when the tube height is less than the tube outer diameter, as is the case for this transducer. 
A BNC cable was soldered to the element electrodes and the actual resonant frequency was obtained by maximizing the transmitted:reflected power ratio, giving a resonant frequency of 444\,kHz. 
All experiments were thus performed at 444\,kHz, without electrical matching. 

\begin{figure}
    \centering
    \includegraphics[width = \columnwidth]{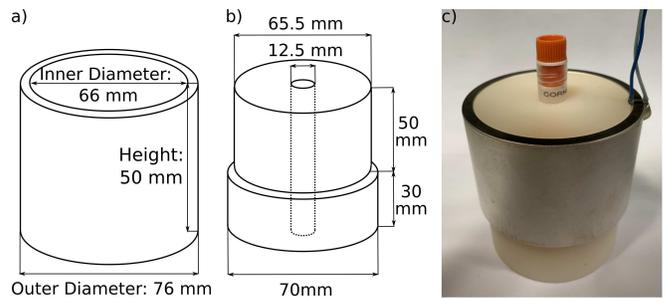}
    \caption{The rewarming setup. a) Transducer dimensions. b) Nylon 6.6 mount dimensions. c) The mounted transducer and a cryovial half-inserted into the nylon 6.6 mount.}
    \label{fig:Tx_Nylon}
\end{figure}

A 71\,mm diameter natural nylon 6.6 rod was machined to fit inside the tubular transducer with a 0.25\,mm tolerance (See Fig. \ref{fig:Tx_Nylon}).
The 0.25\,mm radial tolerance allows for a sliding fit and avoids damaging the transducer via thermal expansion of the nylon 6.6 rod (coefficient of linear thermal expansion: 80 $\mu$m/(m$\cdot$K)) and thermal contraction of the transducer (coefficient of linear thermal expansion -4 to -6 $\mu$m/(m$\cdot$K) in the radial direction \cite{piceramic}). 
The 0.25\,mm gap was filled with deionized water to couple the transducer to the nylon 6.6 rod.
The gap allowed for thermal expansion of the nylon 6.6 and transducer thermal contraction (up to 80$^{\circ}$C), or swelling of the nylon 6.6 due to water absorption. 
A 12.5\,mm diameter hole was machined in the center of the rod for the placement of a cryovial.
Air-backing the tubular transducer further increases the heating efficiency by directing the acoustic energy toward the cryovial and forming a resonant cavity.

\subsection{Biological Medium}

\begin{figure}
    \centering
    \includegraphics[width = \columnwidth]{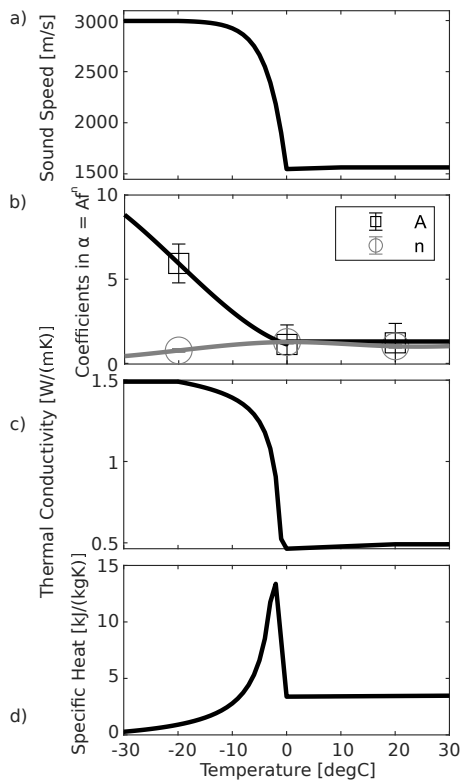}
    \caption{Acoustic and thermal properties of beef as a function of temperature: a) sound speed\cite{miles1974changes}, b) attenuation\cite{shore1986attenuation} fit with splines,  c) thermal conductivity \cite{barrera1983thermal}, and d) specific heat\cite{tavman2007apparent}.}
    \label{fig:BeefAcousticThermalProps}
\end{figure}

Ground beef was chosen as the biological medium to be rewarmed, as the thermal and acoustic properties have been well characterized in work published in the food sciences literature for temperatures ranging from -20$^{\circ}$C to 20$^{\circ}$C \cite{barrera1983thermal,miles1974changes,shore1986attenuation,tavman2007apparent}.
The temperature-driven changes in beef sound speed\cite{miles1974changes}, attenuation\cite{shore1986attenuation}, thermal conductivity\cite{barrera1983thermal}, and specific heat\cite{tavman2007apparent} are displayed in Fig. \ref{fig:BeefAcousticThermalProps}.
Beef has also been used in prior work in ultrasonic thawing, allowing for comparisons of rewarming rates\cite{miles1999high}.
Ground beef was used to increase the ease of handling and shaping.

The changes in beef acoustic and thermal properties over this temperature range present challenges and opportunities for ultrasound-based rewarming.
Thermal conductance decreases threefold as the beef thaws, reducing the diffusion of hot spots.
The sound speed decreases by a factor of nearly two, changing the acoustic impedance mismatch between the beef and the coupling medium.
The attenuation decreases several-fold, reducing the volume rate of heat deposition in thawed samples (this is advantageous for avoiding thermal runaway, but still reduces mean heating rate of a sample).
The specific heat curve shown in Fig. \ref{fig:BeefAcousticThermalProps} incorporates the latent heat of fusion and increases by over an order of magnitude as the beef temperature increases towards zero, reducing the heating rate but potentially increasing the homogeneity of heating within the frozen beef.

\subsection{Coupled acoustic-thermal simulations}

Acoustic and thermal simulations were performed to help design and predict the rewarming behaviour of the experimental setup.
Continuous wave ultrasound propagation was simulated using the k-Wave fluid 2D model (\texttt{kspaceFirstOrder2D}) from version 1.3 of the open-source k-Wave toolbox \cite{treeby2010k,treeby2012modeling} and using Matlab 2019a. 
Rewarming was simulated using the k-Wave \texttt{kWaveDiffusion} code, which solves the Pennes Bioheat Equation\cite{pennes1948analysis,treeby2015contribution}.
Thermal conduction was included in the thermal model, allowing heat to flow from the nylon 6.6 to the beef.
The simulation domain was created using the \texttt{makeDisc} function to generate masks of the nylon 6.6 holder and the frozen beef, and the \texttt{makeCircle} function to generate the tubular source.
The initial temperature distribution was set as 20$^{\circ}$C in the surrounding air and nylon 6.6 and -20$^{\circ}$C in the frozen beef.
The output metrics of the simulation were the spatial compressional pressure profiles in the simulation domain in Pa and the corresponding volume rates of heat deposition $Q$ in W/m$^3$.
$Q$ was calculated using the plane wave assumption, i.e. $Q = \alpha p^2/(\rho c)$ , where $\alpha$ is compressional attenuation in Np/m, $p$ is pressure in Pa, $\rho$ is density in kg/m$^3$, and $c$ is compressional sound speed in m/s.

The acoustic and thermal parameters of frozen beef, nylon 6.6, and air used in simulation are listed in Table \ref{tbl:MediumProps}.
The acoustic and thermal properties of beef displayed in Table \ref{tbl:MediumProps} are those at -20$^{\circ}$C.
The compressional sound speed and attenuation in the nylon 6.6 was measured experimentally at 20$^{\circ}$C.
The acoustic and thermal properties of room temperature air were obtained from the Engineering Toolbox website\cite{ETB}.
The density of air was multiplied by 100 to reduce the computation grid requirements while maintaining effectively the same interfacial behaviour as expected with the physical air density value.

\begin{table}
\caption{\label{tbl:MediumProps}The acoustic and thermal properties (density $\rho$, sound speed $c$, attenuation coefficient $\alpha$, specific heat $C$, and thermal conductivity $k$) of beef at -20$^{\circ}$C and nylon 6.6 and air at room temperature (500\,kHz).}
\begin{ruledtabular}
\begin{tabular}{ccccc}
Property                        &  Beef                      & Nylon 6.6               & Air \\
\hline
$\rho$ [kg/m$^3$]               & 1073 \cite{barrera1983thermal}   & 1120 \cite{Onda}  & 116* \\
$c$ [m/s]                       & 2996 \cite{miles1974changes}     & 2586        & 343 \cite{ETB}\\
$\alpha$  [dB/(cm$\cdot$MHz)]         & 1.21 \cite{shore1986attenuation} & 0.9       & 0\\   
$C$ [J/(kg$\cdot$K)]                  & 902 \cite{barrera1983thermal}    & 1700 \cite{ETB}    & 1000 \cite{ETB}   \\
$k$ [W/(m$\cdot$K)]                   & 1.4908 \cite{tavman2007apparent} & 0.25 \cite{ETB}   & 0.03  \cite{ETB}  \\
\end{tabular}
\end{ruledtabular}
\end{table}

The accurate representation of changes in acoustic and thermal properties of the beef with rewarming requires coupling of acoustic and thermal distributions in simulation.
This was approximated with consecutive acoustic $\rightarrow$ thermal $\rightarrow$ acoustic ... simulations with an update frequency `f' of the acoustic properties and fields and thermal properties.
The change in beef density was small relative to the changes in other acoustic and thermal properties and was consequently fixed at the values reported in Table \ref{tbl:MediumProps}.
The model geometry includes the air surrounding the transducer and coupling medium but convection and air flow around the device are not modelled.
This model should be an acceptable approximation of the rewarming experiments, as the heat transfer coefficients between the nylon 6.6 holder and the surrounding air is low.

The coupled acoustic and thermal simulations were performed in 2D, as a horizontal cross-section of the experimental setup.
The grid size for the 2D simulations was 432 x 432 (83.4324\,mm x 83.4324\,mm, dx =  0.193\,mm), and the acoustic simulations were run for 344\,$\mu$s of acoustic wave propagation (5.6\,ns time step, CFL = 0.01, 20 grid point perfectly matched layer). 
The computation time for one acoustic simulation was approximately 30\,s using an NVIDIA GeForce Titan X GPU in an Intel(R) Xeon(R) CPU E5-2650 0 @ 2.00GHZ server and the \texttt{kspaceFirstOrder2DG} k-Wave code.
The grid size for the 2D rewarming steps was the same (432 x 432), the temporal discretization was 1\,ms, and simulating 200 time steps took approximately 2\,s using the \texttt{kWaveDiffusion} code on the server.
The total computation time, including pre- and post-processing steps, for 100\,s of simulated ultrasonic rewarming was 7.7 hours.
In the thermal simulations, the initial temperature distribution was generated by setting the ground beef temperature to -20$^{\circ}$C and the surrounding nylon 6.6 and air temperature to 20$^{\circ}$C.
This ensured that thermal conduction-based rewarming was incorporated into the model, mimicking the experimental rewarming conditions.
During the first 8 seconds of simulated rewarming, $Q$ was set to zero, representing the time taken to transfer the cryovials from the freezer to the rewarming setup and increase the acoustic power of the transducer to the desired value.

\begin{figure}
    \centering
    \includegraphics[width = \columnwidth]{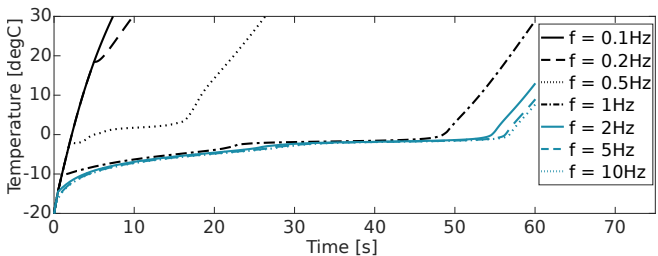}
    \caption{The simulated heating rate decreases as the simulation update frequency (f) increases. Convergence to under a degree difference per time point occurs at f = 5\,Hz.}
    \label{fig:updateconvergence}
\end{figure}

A set of coupled ultrasound and thermal simulations were performed, where the acoustic and thermal properties and fields were updated sequentially.
The update frequency of acoustic and thermal properties and acoustic field was 5\,Hz, obtained via a convergence study that found a $<$1.5$^{\circ}$C/s maximum absolute difference relative to a further doubling of the update frequency.
Figure \ref{fig:updateconvergence} shows the effect of the update rate on temperature evolution: fixing the acoustic and thermal properties at the values shown in Table \ref{tbl:MediumProps} for 10\,s (f = 0.1\,Hz) of heating resulted in an average rewarming rate nearly 400$^{\circ}$C per minute, an over-prediction of realistic rewarming rates.
Further increases in the field update frequency may be necessary to accurately simulate faster rewarming and regions with larger thermal gradients.

\subsection{3D axisymmetric simulations}

The 3D axisymmetric code (\texttt{kspaceFirstOrderAS}) was used to estimate the spatial pressure magnitude distributions\cite{treeby2020nonlinear}.
The simulation domain for the 3D axisymmetric simulation was created using a 2D plane and 2D masks (See Fig. \ref{fig:AxisymmetricFields}) and includes the beef, radially surrounded by the nylon 6.6 holder, which was vertically capped and backed by air.
The source was created using the \texttt{makeLine} function, which becomes a tube when the 2D plane is transformed to axisymmetric coordinates.
Due to the computational cost, the 3D code was not used to perform consecutive acoustic and thermal rewarming simulations.
Running the axisymmetric code with the same CFL number and spatial discretization as the 2D simulations (input grid 324 x 243, 62.5743\,mm by 46.9307\,mm) takes approximately 50\,minutes.
With a similar increase in the computation time for the 3D thermal simulation,  full 3D acoustic and thermal simulations of 100\,s of experimental rewarming would take close to 800 hours. 

\subsection{Experiments}

The acoustic and rewarming capabilities of the tubular transducer were evaluated experimentally.
The acoustic fields generated by the tubular transducer were characterized using hydrophone line and plane measurements, while the rewarming rates were characterized with thermocouples embedded in beef samples.
All analysis was performed in Matlab 2019a.
 
\subsubsection{Acoustic Characterization Experiments}

A set of ultrasound field characterization measurements were performed for the tubular transducer and nylon 6.6 holder.
The tubular transducer was coupled to the nylon holder with deionized water and sealed at the base of the connection between the transducer and nylon 6.6 holder.
The 12.5\,mm nylon 6.6 tube hole was also filled with deionized water for the field characterization measurements.
The ultrasound field was measured inside the tube using a tapered fibre-optic hydrophone (Precision Acoustics, Dorchester UK) mounted to the 3-axis scanning arm of a UMS automated scanning tank (Precision Acoustics, Dorchester UK).
The tapered fibre-optic hydrophone is small (diameter of approximately 120$\mu$m and active element diameter of 10$\mu$m) and the sensitivity is nearly invariant with respect to incident wave angle for frequencies below 1\,MHz, so a correction for the 90$^{\circ}$ incidence angle was not made\cite{morris2005development}.
The transducer was driven with a pulsed (2 cycle) or quasi-continuous signal (100 cycles) generated by a Keysight 33500B Series Waveform Generator and amplified by an  E\&I 1020L, 200\,W amplifier to 50\,V (pulse repetition frequency: 250\,Hz).
The fibre-optic hydrophone voltages (32 averages, 50\,Ms/s, 5000 samples) were digitized with a Tektronix DPO5034B oscilloscope and recorded.
The UMS system was used to perform 1D scans (axial scan dimensions: 40\,mm range, 0.2\,mm steps) and 2D scans (axial dimensions: 20\,mm, 0.5\,mm steps and lateral dimensions: 6\,mm range, 0.2\,mm steps) in the nylon 6.6 tube.
The 1D and 2D scan positions are shown relative to the nylon 6.6 tube in Fig. \ref{fig:EXPfieldprofile}d).
1D scans were repeated across the maximum pressure location with increasing driving voltages (25\,V to 135\,V in approximately 18\,V increments) to generate a pressure vs. voltage calibration.
The fibre-optic hydrophone was calibrated via comparison with a calibrated 0.2\,mm needle hydrophone using measurements of the field of a 270\,kHz source and a 400\,kHz source, and linearly extrapolating to 444\,kHz.
Voltage waveforms were then converted to pressure using this calibration.
Fibre-optic hydrophone measurements were not feasible at higher driving voltages and pressures; at higher pressures the optical fibre flexibility allowed the whole fibre to vibrate with the pulse-repetition frequency of the transducer, distorting the recorded waveforms by changing the position of the active element of the fibre optic hydrophone.

\subsubsection{Rewarming Experiments}

Rewarming experiments were performed with ground beef in 2\,mL cryovials.
Type T thermocouples were used to record the temperature during rewarming experiments (5SRTC-TT-TI-40-1M, Omega, Norwalk, CT, USA).
The thermocouples had a conductor diameter of 0.08\,mm, a bead diameter of approximately 0.24\,mm, and a cable outer diameter of 0.4 x 0.7\,mm.
The thermocouple cable insulation was stripped to expose the conductors, reducing the thermocouple wire diameter exposed to the ultrasound field.
The thermocouple wire diameter was approximately 1/40th of the source wavelength in frozen beef and was sufficiently small relative to the source wavelength to avoid heating artifacts by minimizing the scattering cross-section of the thermocouple and by minimizing the heat capacity and consequent thermal lag of the thermocouple \cite{fry1954determination}
The thermocouple may generate a viscous heating artifact when the ground beef is in a thawed state \cite{fry1954determination,morris2008investigation}.
However, thermocouple-based viscous heating is unlikely to contribute to rewarming in a frozen biological medium.
Ice is much more viscous than soft tissue (at least nine orders of magnitude)\cite{fowler1997glaciers} and calculating the viscous heating artifact for an ultrasound field intercepting a perpendicular thermocouple embedded in ice gives a negligible rate of heat deposition\cite{fry1954determination,tiennot2019numerical}, however, the viscous heating artefact has not been widely explored in solid materials.  

\begin{figure}
    \centering
    \includegraphics[width = \columnwidth]{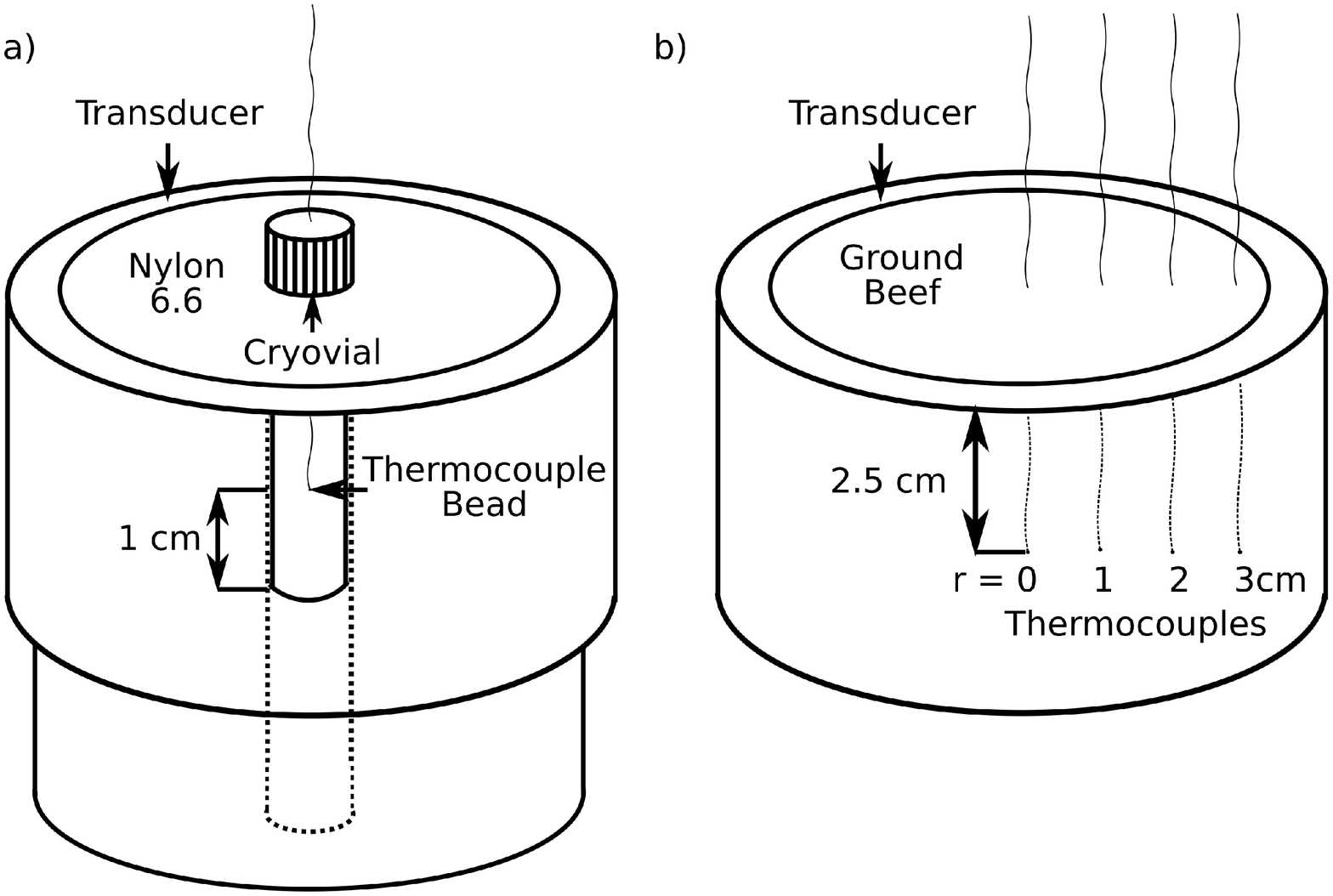}
    \caption{Thermocouple placement a) in the cryovials and b) in the ground beef-filled transducer (r denotes the distance from the center of the transducer).}
    \label{fig:ThermocouplePlacement}
\end{figure}

Corning 2\,mL non-self standing cryovials were used for the rewarming experiments.
Thermocouples were strung between holes punched at the centers of the bottoms of the cryovials and the cryovial caps.
Thermocouple measurements were recorded at 1\,Hz using an NI-TC01 (National Instruments) temperature logger, with the recording starting prior to cryovial loading into the rewarming setup.
Ground British beef (5\% fat content) was loaded into the cryovials and around the thermocouple wires.
The thermocouple beads were pulled back 1\,cm into the cryovial, embedding the bead in the beef (Fig. \ref{fig:ThermocouplePlacement}).
The filled cryovials and embedded thermocouples were then loaded into a -20$^{\circ}$C freezer (LEC-medical) and left to freeze overnight.

The cryovials were loaded individually into the rewarming setup.
The transducer was then driven with a 220\,V amplitude continuous wave signal, corresponding to a 160\,W output from the amplifier.
The signal amplitude at the signal generator was adjusted incrementally to maintain the amplifier power output as the electrical impedance of the transducer changed due to transducer warming. 
The nylon 6.6 holder was cooled in room-temperature water for at least 10 minutes before each experiment to minimize differences in rewarming due to thermal conduction.

Two additional rewarming experiments were performed to demonstrate ultrasonic rewarming with negligible thermal-conduction based heating. 
One experiment was performed with the transducer fully filled with ground beef.
Four thermocouples were embedded, one at the center of the transducer, and the other three spaced 1\,cm apart, extending radially from the first thermocouple (Fig. \ref{fig:ThermocouplePlacement}b). 
The four thermocouples were connected to a TC-08 datalogger (Pico Technologies, Corby, UK), connected to a workstation for temperature logging.
The transducer ground-beef thermocouple assembly was frozen overnight in the -20$^{\circ}$C freezer.
The assembly was then removed from the freezer and rewarmed using the same heating protocol that was used for the cryovial experiments.
A second experiment was performed with a large volume nylon 6.6 holder, identical to the holder used for the cryovials but with a hole with an internal diameter of 46\,mm and a depth of 50\,mm in order to isolate the transducer from the ground beef and replicate the interfaces of the cryovial experiments.
The large volume holder was filled with ground beef (total volume of approximately 80\,mm$^3$), and two thermocouples were embedded (one at the center and one placed 6\,mm radially).
The nylon - ground beef assembly was also frozen overnight in the -20$^{\circ}$C freezer, and in this case the transducer was left at room temperature.

\section{Results}

\subsection{Simulations of the acoustic field in the nylon tube holder}

\begin{figure}
    \centering
    \includegraphics[width = \columnwidth]{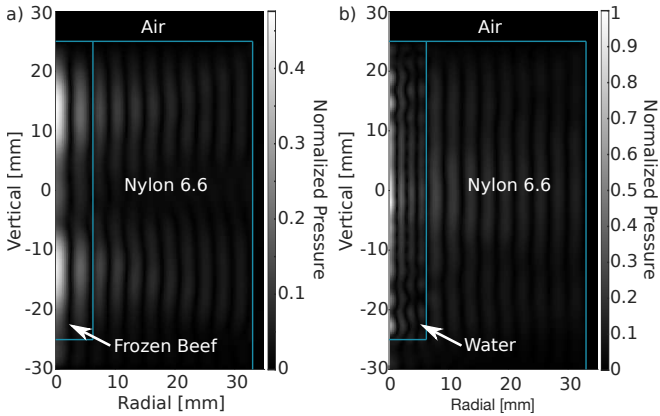}
    \caption{Continuous-wave fields in the nylon 6.6 tube holder and a) the tube filled with frozen beef, or b) the tube filled with water. Magnitudes normalized by the water maximum.}
    \label{fig:AxisymmetricFields}
\end{figure}

Simulations of the experimental setup were performed with the k-Wave axisymmetric code to generate 2D slices of the 3D pressure distribution.
The continuous-wave fields are shown for frozen beef (Fig.\ref{fig:AxisymmetricFields}a) and water (Fig.\ref{fig:AxisymmetricFields}b) in the nylon 6.6 tube.
The simulations show that the source generates a centered focus with a radial distribution of lobes, with some vertical variation in the pressure distribution in the beef but much more in the water distribution.
The water pressure field varies axially due to the low attenuation of ultrasound in water, giving the ultrasound waves longer to interact with the tube boundary conditions and develop axial standing wave patterns.
The combination of standing waves in the water pressure distribution and the higher attenuation of frozen beef results in higher pressures in the water simulation; the maximum pressure in water in Fig. \ref{fig:AxisymmetricFields} is 2.1 times the maximum pressure in the equivalent frozen beef simulation.

\subsection{Experiments}

\subsubsection{Transducer Characterization}

\begin{figure}
    \centering
    \includegraphics[width = \columnwidth]{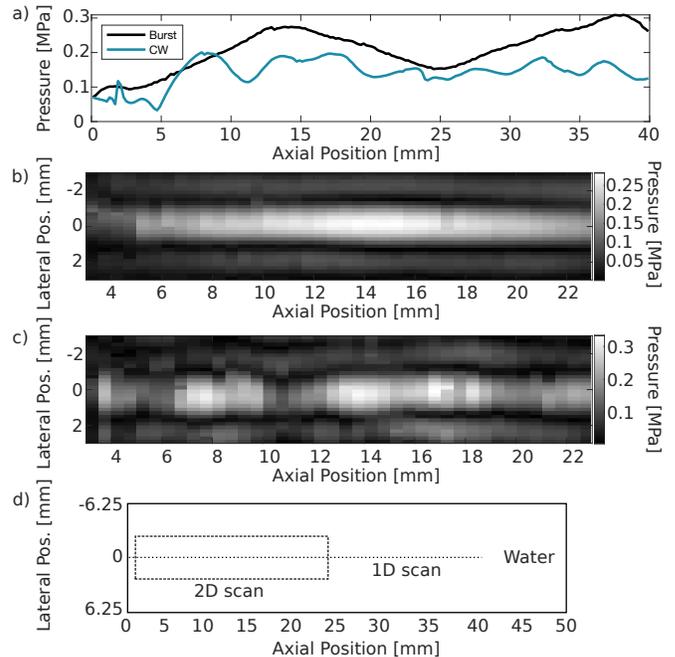}
    \caption{a) Measured axial field profile at the geometric center of the tube, with the transducer driven with a short burst or with a quasi-continuous wave (CW). Measured 2D field profiles for a b) short burst signal and c) quasi-CW signal. d) 1D and 2D scan positions in the water-filled tube.}
    \label{fig:EXPfieldprofile}
\end{figure}

Pressure fields were generated by the transducer with either short bursts or pseudo-continuous wave (CW) signals.
Figure \ref{fig:EXPfieldprofile}a) shows 1D measurements of the burst and CW conditions, taken along the line depicted in Fig. \ref{fig:EXPfieldprofile}d).   
Figure \ref{fig:EXPfieldprofile}b) and c) show 2D measurements of the burst and CW conditions, respectively.
The 2D measurement planes are shown in Fig. \ref{fig:EXPfieldprofile}d). 

\begin{figure}[ht]
    \centering
    \includegraphics[width = \columnwidth]{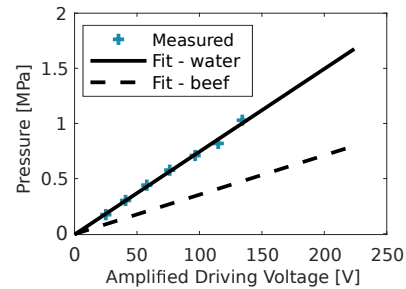}
    \caption{Amplified driving voltage versus pressure in water, extrapolated using simulation to frozen beef.}
    \label{fig:Voltage_vs_Pressure}
\end{figure}

Figure \ref{fig:Voltage_vs_Pressure} shows the relationship between the amplified driving voltage and the maximum recorded pressure in the nylon 6.6 tube. 
Figure \ref{fig:Voltage_vs_Pressure} extends the linear fits to higher driving voltages corresponding to those used in the rewarming experiments; the fibre-optic hydrophone was unable to make accurate pressure measurements at these higher driving voltages because the hydrophone began resonating in the acoustic field.
The simulated water:frozen beef pressure magnitude ratio (2.1) was used to estimate the maximum pressure magnitude in frozen beef for a given driving voltage.
The estimated pressure amplitude at the driving voltage used for the rewarming experiments ($\approx$220\,V) was approximately 0.8\,MPa.
Assuming a linear signal, the mechanical index $in$ $situ$ is given by $MI = 0.8 / \sqrt{0.444} = 1.2$, below the threshold of 1.9 necessary to avoid cavitation in the presence of microbubbles.
This $in$ $situ$ mechanical index indicates that the sonications should not result in mechanical damage to the biological medium encased in the cryovial.

\subsubsection{Rewarming Experiments}

\begin{figure}
    \centering
    \includegraphics[width = \columnwidth]{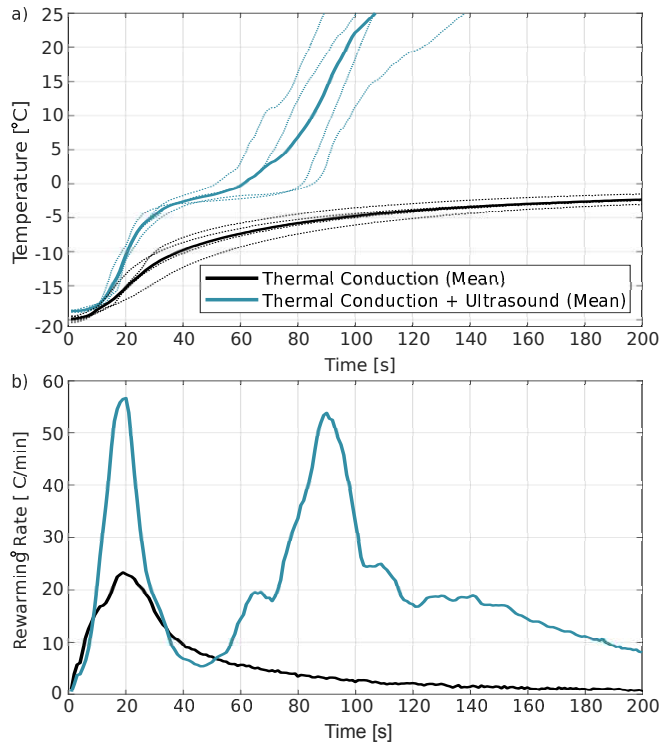}
    \caption{a) Rewarming temperature rise versus time with thermal conduction-only or ultrasound + thermal conduction. b) Heating rates versus time for the two types of experiments. }
    \label{fig:RewarmingEXPs}
\end{figure}

The cryovial temperatures for the four thermal conduction-only rewarming trials and the four ultrasound + thermal conduction rewarming trials are shown in Fig. \ref{fig:RewarmingEXPs}a), along with the mean temperatures for each experiment type. 
The rewarming rates were obtained by taking temperature time derivatives of the mean temperatures and are displayed in Fig. \ref{fig:RewarmingEXPs}b).
Rewarming with ultrasound was at least twice as fast and on average over 10 times faster than with thermal conduction alone, despite the relatively fast rewarming induced by the large temperature gradient between the cryovial and nylon 6.6 holder. 
The thawing time (defined as the time to rewarm from -2$^{\circ}$C to 2$^{\circ}$C) was over an order of magnitude faster when ultrasound was applied. 
Rewarming rates are reported for the thermal conduction-only and thermal conduction + ultrasound experiments in Table \ref{tbl:RewarmingRates}.

\begin{table}
\caption{\label{tbl:RewarmingRates}Rates of temperature rise between -20$^{\circ}$C and 5$^{\circ}$C for cryovial rewarming experiments.}
\begin{ruledtabular}
\begin{tabular}{ccc}
                    & Thermal             & Thermal Cond. \\
                    & Conduction          & + Ultrasound \\
\hline
Max. Heating Rate   & 23$^{\circ}$C/min   & 57$^{\circ}$C/min           \\
Mean Heating Rate   & 2$^{\circ}$C/min    & 23$^{\circ}$C/min           \\
Min. Heating Rate   & 0$^{\circ}$C/min    & 5$^{\circ}$C/min           \\
Thaw time (-2°C to 2°C) & 290$\pm$30\,s     & 24$\pm$2\,s               \\
\end{tabular}
\end{ruledtabular}
\end{table}

\begin{figure}
    \centering
    \includegraphics[width = \columnwidth]{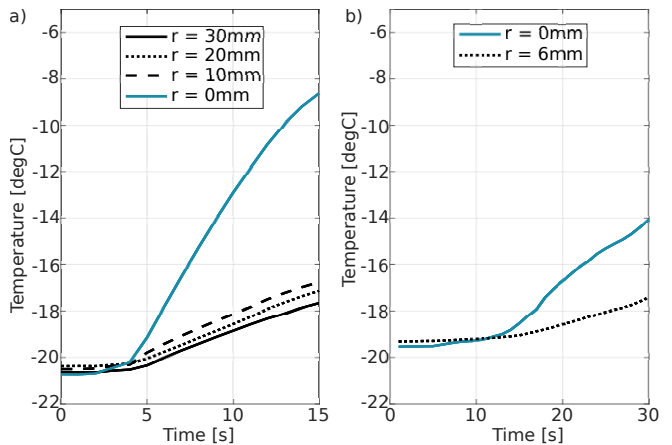}
    \caption{The rewarming curves for a) four thermocouples embedded in the transducer-frozen beef assembly, and b) two thermocouples embedded in the large bore nylon 6.6 tube-beef assembly (thermocouples at radii r from the tube center).}
    \label{fig:BurgerRewarming}
\end{figure}

An additional rewarming experiment was performed with the tubular transducer filled with ground beef and thermocouples (Fig. \ref{fig:ThermocouplePlacement}).
The four thermocouple temperature recordings for this experiment are shown in Fig. \ref{fig:BurgerRewarming}a).
The rewarming rate was highest at the central thermocouple, which averages a rewarming rate of 50$^{\circ}$C per minute from 4\,s to 15\,s.
The approximately 4\,s delay corresponds to the time it took to turn on the source and increase the driving voltage to 220\,V. 
The average rewarming rate at the central thermocouple was faster than the average rewarming rate seen in the cryovial experiments for the same initial rewarming period.
The faster rewarming rate may be due to fewer interfaces reflecting sound away from the center of the large volume beef sample.
This experiment demonstrates ultrasound-based rewarming with minimal influence from thermal conduction and demonstrates the absorption of ultrasonic energy into the frozen beef.

The experiment was repeated with the transducer coupled to a large bore nylon 6.6 tube containing 80\,mL of ground beef. 
The temperature recordings from a central thermocouple and a thermocouple placed 6\,mm radially are shown in Fig. \ref{fig:BurgerRewarming}. 
Similar behaviour is shown in both Fig. \ref{fig:BurgerRewarming}a) and b); the central thermocouple rewarms faster than the thermocouple placed radially.
The sub-zero rewarming rate at the central thermocouple was approximately 3 times slower slower than the central thermocouple in the transducer-frozen beef assembly rewarming experiment (which eliminated several reflective interfaces), and approximately twice as slow as in the cryovials, where thermal conduction sped up the early rewarming.

\subsection{Comparison of simulated and experimental rewarming}

\begin{figure}
    \centering
    \includegraphics[width = \columnwidth]{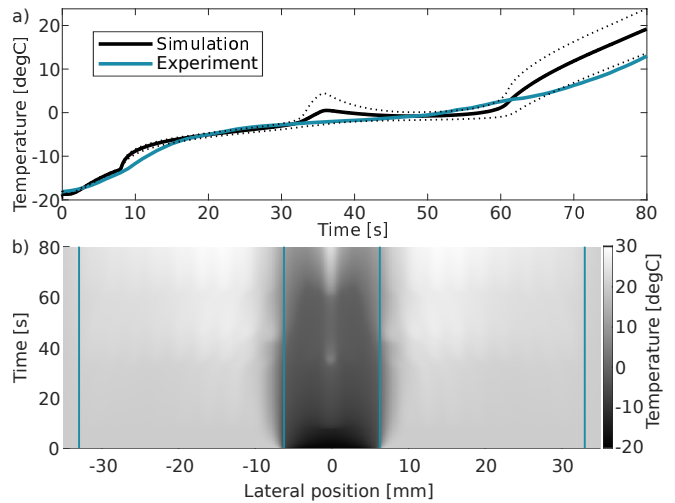}
    \caption{a) experimental vs. simulated rewarming (max., mean, and min. in a 1\,mm$^2$ area centered in the tube) b) 1D slices through the simulated cryovial for 80\,s of rewarming. The nylon 6.6 and beef edges are depicted with blue lines.}
    \label{fig:USsim_vs_USexp}
\end{figure}

Rewarming was simulated with a simplified 2D model and compared with the experimental temperature recordings, as shown in Fig. \ref{fig:USsim_vs_USexp}a).
The simulation replicates the experimental rewarming below 0$^{\circ}$C accurately, but the behaviour of the simulation above 0$^{\circ}$C deviates from the experimental measurements. 
Figure \ref{fig:USsim_vs_USexp}a) shows a brief period at 35\,s where the simulated maximum temperature exceeds 0$^{\circ}$C, then cools again.
In simulation, this fast heating at 0$^{\circ}$C results from a brief but drastic increase in pressure and the corresponding volume rate of heat deposition.
The increase in pressure may be due to the beef sound speed reaching the perfect distribution to create a 444\,kHz resonant chamber.
This effect is not seen in experiment, possibly due to the assumptions and simplifications made in the simulation.
The simulation does not incorporate ground beef inhomogeneity, the acoustic properties of the beef around 0$^{\circ}$C may not be accurate, and shear waves that may arise in the experiment are not incorporated into the simulation.
The simulation also shows the beef rewarming to above 0$^{\circ}$C slightly earlier than was measured in experiment. 
This may be the result of using a 2D simulation to reproduce a 3D experiment; the experimental rewarming setup uses a finite length tube, while the 2D simulation assumes the tube length is infinite and that heat only diffuses across the 2D plane, confining the heat in the third dimension.

The temperature slices from the rewarming simulations (Fig. \ref{fig:USsim_vs_USexp}b) show that below 0$^{\circ}$C, the rewarming of the periphery from thermal conduction and the central ultrasonic rewarming act together to reduce the temperature differential in the cryovial.
Before 35\,s, the maximum temperature differential within the cryovial was between 1$^{\circ}$C and 3$^{\circ}$C.
At 35\,s the simulated fast rewarming at the center of the cryovial results in a maximum temperature differential of 7$^{\circ}$C.
After 35\,s, when the mean temperature exceeds 0$^{\circ}$C, the absorption of ultrasound decreases which results in cooling at the centre of the cryovial and further rewarming towards the edges.
After 60\,s, the entire cryovial has thawed and the center of the cryovial again rewarms faster than its surroundings.

\section{Discussion}

The work presented here experimentally demonstrates the principle of ultrasonic rewarming and provides a foundation for further investigation and development of the technique. 
Ultrasonic rewarming of frozen ground beef was demonstrated in 2\,mL cryovials at average rates over ten times faster than with thermal conduction alone.
The ultrasound system developed in this work to rapidly rewarm 2\,mL cryovials may be a useful tool to study the recovery of cryopreserved cells commonly stored in these sterile containers. 
The transducer and nylon 6.6 setup was characterized with fibre-optic hydrophone measurements in the nylon tube cavity. 
Simulation was then used to estimate the water to frozen beef pressure ratio, and it was estimated that the pressure amplitude at the centre of the cryovial during rapid rewarming of the beef was approximately 0.8\,MPa. 
Ultrasonic rewarming was also demonstrated in two larger beef volumes where thermal conduction from the surroundings was minimized and faster ultrasonic rewarming was demonstrated at the transducer focus than at radial positions. 
A hypothesized advantage of ultrasonic rewarming over radiofrequency-based methods is the decrease in absorption as the medium thaws, reducing the hot spot problem that can occur with radiofrequency-based methods. 
The simulations in this work support this, showing a small temperature differential across the cryovial during the transition through sub-zero temperatures; this will be confirmed by future experimental work.

Prior work in ultrasonic rewarming has been limited to long timescales (tens of minutes to hours)\cite{miles1999high} or to simulations of rewarming below -20$^{\circ}$C that neglected the temperature-dependence of the thermal and acoustic parameters of biological tissues\cite{olmo2020use}.
This work experimentally demonstrates fast ultrasonic rewarming at high sub-zero temperatures, through the transition zone from frozen to thawed.
The mean rewarming rate measured at the centre of the cryovials was 23$^{\circ}$C per minute, and simulations suggest that the temperature gradient across the sample is of the order of a few degrees. 
For slow frozen materials, slow warming rates of the order of 1$^{\circ}$C per minute are feasible, but increases in post thaw viability have been demonstrated where small volumes cell samples were warmed at faster rates in a water bath ($\sim$15$^{\circ}$C/min vs 0.6$^{\circ}$C/min)  \cite{kilbride2017cryopreservation}. 
The faster warming rates measured here may further improve cell viability in these type of materials.
Previously, larger volumes of several litres have been rewarmed slowly (0.6$^{\circ}$C per minute) to reduce temperature differentials with acceptable 72\,h post-rewarming cell viability\cite{kilbride2017cryopreservation}. The addition of ultrasound could increase the rewarming rate at the center of these volumes, improving the bulk rewarming rate without the risk of generating a large temperature differential within the volume, consequently reducing the time needed for the cells to regain normal functionality.

For successful rewarming of vitrified media, warming rates of 100$^{\circ}$C/min or more (an order of magnitude faster than critical cooling rates) are commonly cited as necessary to avoid ice nucleation and growth\cite{olmo2020use,robinson2002electromagnetic,fahy2015principles}. However, this may be an underestimate and the critical rewarming rate may be multiple orders of magnitude faster than the critical cooling rate\cite{baudot2004thermal,fahy1987biological,fahy2015principles}.
With further optimisation of the ultrasonic warming device, it may be possible to increase rewarming rates towards the 50$^{\circ}$C/min generated by magnetic nanoparticle heating or further towards 100$^{\circ}$C/min which may be in the lower limit of rates required to warm vitrified materials.

To facilitate faster warming, transducer acoustic power output will be increased by optimising the efficiency of the source e.g. by electrical impedance matching and addition of an acoustic matching layer between the transducer and the nylon 6.6 holder.
The transducer frequency used here was chosen by balancing the focal dimensions with attenuation and lies within the 0.4-1\,MHz range of ultrasound frequencies previously found to be capable of slow ultrasound-based thawing\cite{miles1999high}. Together with increased knowledge of the frequency dependence of absorption in frozen materials, further optimisation of the source frequency may increase warming rates or ensure uniform heating of larger volumes. 
The time course of energy delivery could also be adjusted to deliver a more constant rate of warming through the temperature range, reduce temperature gradients and reduce warming above 0$^{\circ}$C.
The use of nylon 6.6 to couple the transducer to the cryovials generated fast rewarming rates, but a deformable or conformal coupling layer may be needed to rewarm larger volumes where there are significant shape or volume changes during rewarming. 
The tubular transducer geometry is well-matched to the cryovial geometry, but larger-volume ultrasonic rewarming will require focal steering achieved though the development of a phased array transducer\cite{olmo2020use}.

The relative contributions of scattering and absorption to attenuation in frozen biological materials are not known, but the experiments presented here demonstrate that there is some contribution of absorption to the attenuation that results in warming of the samples.
The rewarming simulations were performed with the acoustic absorption coefficient set to the measured attenuation coefficient of beef. 
This is unlikely to represent the true level of absorption although the the simulation matches the experiment well. 
This may be partly because the rewarming setup is effectively a closed acoustic system meaning that scattered sound will eventually be absorbed. 
Further investigation of the contributions of absorption and scattering to acoustic attenuation of frozen biological materials is required for the further development of this technology. 
Further characterisation of the acoustic and thermal properties of different cell and tissue types at sub-zero temperatures, as well as their dependence on cooling rate and mechanism (e.g. progressive solidification, network solidification, vitrification), and presence of cryoprotective and ice nucleating agents is also critical in understanding and optimising the delivery of energy for rewarming. 
This information will enable more accurate modelling for development of more sophisticated ultrasonic rewarming platforms and optimisation of rewarming rates.

Ultrasonic rewarming research is in its early stages and there are many remaining parameters to investigate.
Rewarming of different samples of greater clinical relevance will be investigated, for example by examining rewarming rates in different cell types and their dependence on cooling rate and cryoprotectant type and concentration.
For example, DMSO, glycerol, and other cryoprotectants are used in concentrations between 10-60\% w/v depending on the cooling method, and are likely to influence the acoustic properties and consequent ultrasonic rewarming rates that can be achieved. 
In future work, warming from the low (-140$^{\circ}$C to -80$^{\circ}$C) temperatures used for long-term cryopreservation will be investigated. It is likely that attenuation and absorption remain relatively constant at lower sub-zero temperatures, suggesting that fast ultrasonic rewarming can also be achieved at over this range.
Cell survival studies will be performed to evaluate the effect of rewarming rate on cell viability which is critical for optimising the rewarming and to investigate any other effects of exposure to ultrasound. 
Lower `power' ultrasonic frequencies (20-100\,kHz) can generate ice nucleation\cite{chow2005study,dalvi2017review}, which when performed without the formation of large ice crystals can improve cell viability \cite{morris2013controlled}. 
Further work is needed to determine whether higher frequency ultrasound generates ice nucleation and whether there is time for the ice crystals to grow during rapid ultrasonic rewarming.

\section{Conclusion}

A custom tubular ultrasonic transducer setup was built to demonstrate ultrasonic rewarming in 2\,mL cryovials.
The rewarming was performed from -20$^{\circ}$C to above 5$^{\circ}$C.
Experiments demonstrated the rewarming of frozen samples with sub-megahertz pressures, at a maximum rate of 57$^{\circ}$C per minute and at an average rate of over ten times faster than with thermal conduction alone.
The experiments demonstrated that absorption forms a part of frozen beef attenuation, and that ultrasound may be a viable method for rewarming cryopreserved samples.
Acoustic and thermal simulations matched well with experiment and suggest warming takes place with low thermal gradients across the sample. 
This work presents a step towards using ultrasound to rewarm samples cryopreserved in cryovials, and develops the knowledge and tools that will be required to rewarm larger cryopreserved volumes.

\section{Acknowledgements}

This work was supported in part by a UKRI Future Leaders Fellowship [grant number MR/T019166/1], and in part by the Wellcome/EPSRC Centre for Interventional and Surgical Sciences (WEISS) (203145Z/16/Z). 
For the purpose of open access, the author has applied a CC BY public copyright licence to any Author Accepted Manuscript version arising from this submission.
The authors wish to thank Professor Barry Fuller and Professor Clare Selden for useful discussions on this work.


\bibliography{references}

\end{document}